\documentclass[conference]{IEEEtran}
\IEEEoverridecommandlockouts
\usepackage{bm}
\usepackage{amsthm}
\usepackage{libertine}
\usepackage[utf8]{inputenc}
\usepackage[margin=1in]{geometry}
\usepackage{amsmath,amssymb}
\usepackage{multicol}
\usepackage[shortlabels]{enumitem}
\usepackage{siunitx}
\usepackage{cancel}
\usepackage{graphicx}
\usepackage{pgfplots}
\usepackage{listings}
\usepackage{tikz}
\usepackage{xcolor}
\usepackage{subfig}
\usepackage{comment}
\usepackage{mdframed}
\usepackage{bm}

\usepackage{hyperref}

\pgfplotsset{width=10cm,compat=1.9}
\usepgfplotslibrary{external}
\allowdisplaybreaks

\newcommand{\RR}{{\mathbb{R}}}
\newcommand{\NN}{{\mathbb{N}}}
\newcommand{\UU}{{\mathbb{U}}}
\newcommand{\YY}{{\mathbb{Y}}}

\newcommand{\KK}{{\mathbb{K}}}
\newcommand{\HH}{{\mathbb{H}}}

\newcommand{\Pf}{{\mathcal{P}}}

\newcommand{\sH}{\mathcal{H}}

\newcommand{\vH}{{\bm{\sH}}}

\newcommand{\calH}{{\mathcal{H}}}

\newcommand{\knl}{\mathfrak{K}}
\newcommand{\Knl}{\mathcal{K}}

\newcommand{\Lcal}{\mathcal{L}}

\newenvironment{Proof}{{\it Proof}:}{$\Box$}

\newcommand{\ipH}[2]{\left\langle#1,#2\right\rangle_\mathcal{H}}
\newcommand{\ipvH}[2]{\left\langle#1,#2\right\rangle_{\vH}}
\newcommand{\ipRn}[2]{\left\langle#1,#2\right\rangle_{\mathbb{R}^n}}

\newcommand{\normvH}[1]{\left \Vert #1\right \Vert _{\pmb{\mathcal{H}}}}
\newcommand{\normRd}[1]{\left \Vert #1\right \Vert _{\YY}}

\newcommand{\normU}[1]{\left \Vert #1\right \Vert _{\mathbb{U}}}

\newtheorem{theorem}{Theorem}

\newtheorem{defn}{Definition}

\title{Vector-Valued Native Space Embedding for \\ Adaptive State Observation}
 
\begin{document}

\author{Shengyuan Niu$^1$, Haoran Wang$^1$, Heejip Moon$^2$, Andrea L'Afflitto$^3$, Andrew Kurdila$^1$, Daniel Stilwell$^2$
\thanks{$^1$S. Niu, H. Wang, and A. Kurdila are with the Department of Mechanical Engineering, Virginia Tech, Blacksburg, VA 24061, USA. Email:
{\tt \{syniu97, haoran9, kurdila\}@vt.edu}.}
\thanks{$^2$H. Moon and D. Stilwell are with the Bradley Department of Electrical \& Computer Engineering, Virginia Tech, Blacksburg, VA 24061, USA. Email:
{\tt \{hmoon5, stilwell\}@vt.edu}.}
\thanks{$^3$A.  L'Afflitto is with the Grado Department of Industrial and Systems Engineering, Virginia Tech, Blacksburg, VA 24061, USA. Email: {\tt a.lafflitto@vt.edu}.}
\thanks{This work is supported in part by the Office of Naval Research (ONR) project number N00014-24-1-2267.} 
}

\maketitle

\begin{abstract}
    This paper combines vector-valued reproducing kernel Hilbert space (vRKHS) embedding with robust adaptive observation, yielding an algorithm that is both non-parametric and robust.
    The main contribution of this paper lies in the ability of the proposed system to estimate the state of a plan model whose matched uncertainties are elements of an infinite-dimensional native space.
    The plant model considered in this paper also suffers from unmatched uncertainties.
    Finally, the measured output is affected by disturbances as well.
    Upper bounds on the state observation error are provided in an analytical form. 
    The proposed theoretical results are applied to the problem of estimating the state of a rigid body.
\end{abstract}

\section{Introduction}
This paper presents an adaptive system able to estimate the state of a nonlinear plant model, whose matched nonlinearities lie in a user-defined infinite-dimensional reproducing kernel Hilbert space (RKHS -- also known as native space) of vector-valued functions.
This system is also designed to be robust to finite-dimensional unmatched uncertainties and disturbances in the measured output.
To the authors' knowledge, this is the first work to propose an adaptive observer in a native space framework and for vector-valued matched uncertainties.

Recently, the authors commenced a systematic effort to port existing robust and adaptive control techniques from a parametric to a non-parametric framework \cite[Ch. 5]{RKHS_MRAC_book}.
Thus far, control techniques such as control Lyapunov functions, backstepping, and model reference adaptive control (MRAC) have been ported to a native space setting \cite{RKHS_paruchuri2020,RKHS_guo2022adaptive,RKHS_guo2020approximations,RKHS_guo2022partial,RKHS_burns2023}. 
Within this framework, this paper is the first to address the observation and estimation problem for multi-input multi-output systems. 
A key advantage in using RKHS embedding methods lies in the ability to counter infinite-dimensional matched uncertainties, whereas existing methods rely on finite-dimensional representations of uncertainties based on a regressor vector or a similar structure.
Furthermore, leveraging native space theory, it is possible to deduce tight upper bounds on the error in approximating matched uncertainties through the proposed approaches \cite[Ch. 3]{RKHS_MRAC_book}.

The proposed theoretical results are applied to a problem of practical interest, namely, the observation of the state of a rigid body.
It is worthwhile recalling that the rigid body model is typically employed to capture the dynamics of uncrewed aerial vehicles (UAVs) \cite{MARSHALL2021390},
autonomous underwater vehicles (AUVs) \cite{njaka2022guide}, and
spacecraft \cite{doi:10.2514/1.G002309} to name a few.
In these applications, however, a characterization of the vehicle, especially of larger vehicles, and of the disturbances acting on them is usually non-trivial.
Numerical methods usually require an intense effort to produce quality results and
experimental methods are usually costly.
Thus, estimation methods applicable to broad classes of uncertainties and that do not rely on a regressor vector or matrix to characterize uncertainties and disturbances are preferable. 

This paper is structured as follows.
In Section \ref{section_RKHS}, we summarize key properties of native spaces and recall essential results needed throughout this work.
In Section \ref{section_Adaptive_Law}, we present a theoretical result that captures the ideal behavior of the proposed adaptive observer.
This observer involves an adaptive law evolving over the native space of functional uncertainties affecting the plant dynamics.
In practice, this adaptive law is not implementable since it is infinite-dimensional.
Thus, in Section \ref{section_implementation}, we present finite-dimensional approximations of the results in Section \ref{section_Adaptive_Law}.
Section \ref{section_numerical_example} demonstrates the applicability of the proposed results through numerical simulations involving the translational and rotational dynamics of a rigid body.
Finally, Section \ref{section_conclusion} draws conclusions and outlines future work directions. 

\section{Notation on Vector-valued Reproducing Kernel Hilbert Space} \label{section_RKHS}
In this section, we discuss elements of RKHS theory.
For a detailed account on this topic, see \cite[Ch. 3]{RKHS_MRAC_book}.
Let $\Omega \subset \YY \triangleq \RR^m$ and $\knl : \Omega \times \Omega \to \RR$ be a Mercer kernel \cite[Def. 3.6]{RKHS_MRAC_book}.
We define the scalar-valued native space
\begin{align}
    \sH\triangleq \overline{\text{span}\{ \knl_y(\cdot) \ : y \in \YY \} },  \label{eqn_ch03_def_sH}
\end{align}
where $\knl_y(\cdot) \triangleq \knl(y,\cdot)$, and the closure is taken with respect to the candidate inner product defined as
\begin{align}
    \ipH{\knl_{y_1}}{\knl_{y_2}}\triangleq \knl(y_1,y_2), \text{ for any }  y_1,y_2\in \YY. \label{eqn_ch03_def_inner_prod_RKHS}
\end{align}
Given $\sH$, we can define the diagonal operator-valued kernel $\Knl : \YY \times \YY \to \Lcal(\UU)$ so that $\Knl(y_1,y_2) \triangleq \knl(y_1,y_2) I_m$ for all $y_1,y_2 \in \YY$, where $I_m \in \RR^{m \times m}$ denotes the identity matrix, and the vector-valued native space $\vH \triangleq \calH^m \triangleq \calH \times \cdots \times \calH$. 
Given $f \in \vH$, we define the \emph{evaluation operator} $E_y : \vH \to \UU \triangleq \RR^m$ so that
\begin{align}
    E_y f = f(y) \in \UU, \text{ for all } f \in \HH \text{ and for all } y \in \Omega,
\end{align}
and we denote the \emph{adjoint of the evaluation operator} as $E_y^{\star} : \vH \to  \UU$.
Both the evaluation operator and its adjoint are bounded and linear.

Given the \emph{set of centers} $\Xi_N \subset \YY$, we consider the finite-dimensional RKHS  
\begin{align}
    \vH_N \triangleq \{\Knl_{\xi_i}(\cdot) \alpha_i : \xi_i\in \Xi_N, \alpha_i\in \UU \}.
\end{align}
to approximate the infinite-dimensional RKHS $\vH$,
where
$\Knl_{\xi_i}(\cdot) \triangleq \Knl(\xi_i,\cdot)$.
Thus, we can decompose any $f \in \vH$ as $f = \Pi_N f+ (I-\Pi_N)f$, where $\Pi_N: \vH \to \vH_N$ denotes the \emph{projection operator} from $\vH$ onto $\vH_N$ \cite[Def. 2.51]{RKHS_MRAC_book}.
The projection error can be bounded as $\normU{E_y (I-\Pi_N) f}\leq  \sup_{y \in \Omega} \bm{\Pf}_N(y) \normvH{f}$ \cite[Th. 3.22]{RKHS_MRAC_book}, where $\bm{\Pf}_N(\cdot)$ denotes the \emph{power function}
\begin{align}
    \bm{\Pf}_N(y) \triangleq \max_{i=1,\ldots,m}&\sqrt{|\Knl_{ii}(y,y)-\Knl_{N,ii}(y,y)|}, \nonumber\\
    &\hspace{7.0em} \text{ for all } y \in \YY, \label{eqn_power_function}
\end{align}
$\Knl_{ii}$ denotes the element of the $i$th row and $i$th column of $\Knl(\cdot,\cdot)$, and $\Knl_N(y_1,y_2) \triangleq \Knl_{\Xi_N}(y_1) \KK_{\Xi_N}^{-1} \Knl_{\Xi_N}(y_2)$ for all $y_1, y_2 \in \Omega$,
\begin{align}
    \Knl_{\Xi_N}(y) \triangleq \left[\Knl(y,\xi_1), \ldots, \Knl(y,\xi_N) \right] \in \RR^{m \times mN}, \label{eqn_kernel_matrix}    
\end{align}
and
\begin{align}
    \KK_{\Xi_N} \triangleq
    \begin{bmatrix}
        \Knl(\xi_1,\xi_1) & \dots & \Knl(\xi_1,\xi_N) \\
        \vdots & \ddots & \vdots \\
        \Knl(\xi_N,\xi_1) & \dots & \Knl(\xi_N,\xi_N)
    \end{bmatrix} \in \RR^{mN \times mN} \label{eqn_grammian_matrix}
\end{align}
denotes the \emph{Grammian matrix}.
\section{Problem Statement}\label{section_problem_statement}
In this paper, we investigate the problem of estimating the state of the nonlinear system
\begin{subequations}\label{eqn_plant_dynamics}
    \begin{align}
        \dot{x}(t) &= A x(t)+ B\left(u(t) + E_{y(t)}f \right), \nonumber \\
        &\hspace{8.0em} x(t_0) = x_0, \quad t \geq t_0,  \label{eqn_plant_dynamics_state} \\
        y(t) &= Cx(t) + \delta(t), \label{eqn_plant_dynamics_output}
    \end{align}
\end{subequations}
where $x : [t_0,\infty) \to \RR^n$ denotes the \emph{state vector}, $u : [t_0,\infty) \to \RR^m$ denotes the \emph{control input}, which is bounded and user-defined, $A \in \RR^{n\times n}$, $B\in \RR^{n\times m}$, $C\in\RR^{m \times n}$, the pair $(A,B)$ is controllable, the pair $(A,C)$ is observable, the linear dynamical system
\begin{subequations}
    \begin{align}
        \dot{x}(t) &= A x(t)+ B u(t), \quad x(t_0) = x_0, \quad t \geq t_0, \\
        y(t) &= Cx(t),
    \end{align}
\end{subequations}
is strictly positive real, the \emph{measurement error} $\delta : [t_0,\infty) \to \UU$ is continuous and bounded, that is, $ \Vert \delta(t) \Vert \leq \overline{\delta}$, and $f \in \vH(\Omega,\UU)$ is unknown.

\section{A Distributed Parameter System Solution}\label{section_Adaptive_Law}
To address the proposed estimation problem, consider the \emph{adaptive observer}
\begin{align}
    \dot{\hat{x}}(t) &= A \hat{x}(t) + L\big(y(t)-C\hat{x}(t)\big) \nonumber \\
    &\quad + B \left( u(t) + E_{y(t)} \hat{f} \right), \quad \hat{x}(t_0) = \hat{x}_0, \quad t \geq t_0,
    \label{eqn_state_estimator}
\end{align}
where $L \in \RR^{n \times m}$ chosen such that $A_e \triangleq A-LC$ is Hurwitz,
the \emph{adaptive gain} $\hat{f} : [t_0,\infty) \times \Omega \to \vH$ denotes the solution of the \emph{adaptive law}
\begin{align}
    \frac{\partial}{\partial t}\hat{f}(t,\cdot) &= \mu_{\rm step}\left(e(t),E_0 \right) \Gamma_f E^*_{y(t)} (y-C\hat{x}), \nonumber \\
    &\hspace{6.0em} \hat{f}(t_0,\cdot) = f_0, \quad t \geq t_0,
    \label{eqn_adaptive_law_infinite_dim}
\end{align}
the \emph{adaptive rate matrix} $\Gamma_f \in \RR^{m \times m}$ is symmetric and positive-definite,
\begin{align}
	\mu_{\rm step}(e,E_0) &\triangleq \left \{ \begin{array}{ll} 1, & \qquad{} \mbox{if } e \notin \mathring{\mathcal{B}}_{E_0}(0_n), \\
			0, & \qquad{} \mbox{otherwise},
		\end{array} 
	\right. \label{eqn_ch07_step_function}
\end{align}
denotes the \emph{step function},
the closed ball $\overline{\mathcal{B}}_{E_0}(0_n)$ centered at $0_n$ and of radius 
\begin{align}
    E_0 \triangleq \frac{\overline{\delta} \Vert L^{\rm T} P \Vert}{\epsilon \lambda_{\min}(P) + \lambda_{\min}(W^{\rm T} W)}
\end{align}
denotes the \emph{dead-zone}, and
the symmetric positive-definite matrix $P \in \RR^{n\times n}$, the matrix $W \in \RR^{p \times n}$, and the scalar $\epsilon > 0$ verify the Lur'e equations
\begin{subequations}
    \begin{align}
    A_e^{\rm T} P + P A_e &= -W^{\rm T} W - \epsilon  P, \\
    PB &= C^{\rm T}.
\end{align}    
\end{subequations}
The effectiveness of the observer \eqref{eqn_state_estimator} and the adaptive law \eqref{eqn_adaptive_law_infinite_dim}
 is captured by the following result.

 \begin{theorem}\label{theorem_DPS}
    Consider the nonlinear dynamics \eqref{eqn_plant_dynamics} and the observer \eqref{eqn_state_estimator}.
    If the dynamical system given by \eqref{eqn_plant_dynamics} and \eqref{eqn_state_estimator} has a unique solution
    $(\hat{x},\tilde{f}) \in \RR^n \times C^1([t_0,\infty),\vH)$ for every initial condition,
    then, there exists $T \geq t_0$ such that $e(t) \in \overline{\mathcal{B}}_{E_0}(0_n)$ for all $t \geq T$.
\end{theorem}

\begin{proof}
    Let $e(t) \triangleq x(t)-\hat{x}(t)$, $t \geq t_0$, denote the \emph{state observation error}.
    It follows from \eqref{eqn_plant_dynamics} and \eqref{eqn_state_estimator} that the state observation error dynamics are given by
    \begin{align}
        \dot{e}(t) &= A_e e(t)+B E_{y(t)} \tilde{f} - L\delta(t), \nonumber \\
        &\hspace{6.0em} e(t_0) = x_0 - \hat{x}_0, \quad t \geq t_0,
        \label{eqn_error_dyn}
    \end{align}
    where $\tilde{f} \triangleq f-\hat{f}$.
    Thus, consider the \emph{Lyapunov function candidate}
    \begin{align}
        V(e,\tilde{f}) &= \ipRn{e}{Pe} + \ipvH{\tilde{f}}{\Gamma_f^{-1}\tilde{f}}, \nonumber \\
        &\hspace{10.0em} (e,\tilde{f}) \in \RR^n \times \vH. \label{eqn_lyapunov_function}
    \end{align}
    Taking the time derivative of \eqref{eqn_lyapunov_function} along the trajectories of \eqref{eqn_error_dyn} and \eqref{eqn_state_estimator}, it holds that
    \begin{align}
        &\dot{V}(t,e,\tilde{f}) \nonumber \\
        &= \ipRn{\dot{e}(t)}{Pe}+\ipRn{e}{P\dot{e}(t)}+2\ipvH{\partial_t \tilde{f}(t,\cdot)}{\Gamma_f^{-1}\tilde{f}} \nonumber \\
        &=\ipRn{A_e e+B(E_{y(t)}\tilde{f})-L\delta(t)}{Pe}\nonumber\\  
        &\quad +\ipRn{e}{PA_e e+PB(E_{y(t)}\tilde{f})-PL\delta(t)}\nonumber\\
        &\quad-2\ipvH{\partial_t \tilde{f}(t,\cdot)}{\Gamma_f^{-1}\tilde{f}} \nonumber \\
        &= \ipRn{e}{\left(A_e ^{\rm T}P + P A_e \right) e}\nonumber\\
        &\quad-2\ipRn{e}{P L \delta(t)} + 2 \ipvH{\tilde{f}}{E_{y(t)}^* B^{\rm T} P e}\nonumber\\
        &\quad \quad -2\ipvH{\partial_t \tilde{f}(t,\cdot)}{\Gamma_f^{-1}\tilde{f}} \nonumber\\
        &= -e^{\rm T}W^{\rm T}We - \epsilon e^{\rm T} P e - 2\ipRn{e}{PL\delta(t)}\nonumber\\
        &\quad + 2\left(\ipvH{\tilde{f}}{E_{y(t)}^*Ce-\Gamma_f^{-1}(\Gamma_f E^*_y (y-C\hat{x})))}\right)\nonumber\\
        &= -e^{\rm T}W^{\rm T}We - \epsilon  e^{\rm T} P e + 2 \overline{\delta} \Vert L^{\rm T} P \Vert \Vert e \Vert, \nonumber \\
        &\hspace{8.0em} (t,e,\tilde{f}) \in [t_0,\infty) \times \RR^n \times \vH.
    \end{align}

    Now, with the adaptive law \eqref{eqn_state_estimator},
	if $e \notin \overline{\mathcal{B}}_{E_0}$, then
	$\dot{V}(t,e,\tilde{f}) < 0$;
	if $e \in \partial \mathcal{B}_{E_0}$, then
	$\dot{V}(t,e,\tilde{f}) = 0$; and
	if $e \in \mathring{\mathcal{B}}_{E_0}$, then
	the sign of 
	$\dot{V}(t,e,\tilde{f})$
	is undefined.
	Therefore,
	if $e(t_0) \notin \overline{\mathcal{B}}_{E_0}$, then
	there exists $t_1 > t_0$ such that
	$e(t) \in \partial \mathcal{B}_{E_0}$.
	The time $t_1$ can be estimated by noting that
	$\dot{V}(t,e,\tilde{f}) \leq - C$, where
	\begin{align*}
		C &\triangleq \sup \bigg \{-\dot{V}(t,e,\tilde{f}): \nonumber \\
        &\quad (t,e,\tilde{f}) \in [t_0,\infty) \times \left(\overline{\mathcal{B}}_{\Vert e(t_0) \Vert} \setminus \mathring{\mathcal{B}}_{E_0} \right) \times \vH \bigg \}
	\end{align*}
	is positive, and, hence,
	\begin{align}
		t_1 \leq t_0 - C^{-1} V(e(t_0),\tilde{f}(t_0)).
	\end{align}
	Since $\dot{V}(t,e(t),\tilde{f}(t))$ is not sign definite for all $t > t_1$ such that
	$e(t) \in \mathring{\mathcal{B}}_{E_0}$ and
	$\dot{V}(t,e(t),\tilde{f}(t)) \leq 0$ for all $t \geq t_1$ such that
	$e(t) \in \partial \mathcal{B}_{E_0}$,
	there exist time sequences
	$\{ t_{2k+1} \}_{k = 0}^M$ and $\{ t_{2k} \}_{k = 1}^M$ 
	with $M \in \overline{N} \cup \{\infty\}$
	such that
	$e(t) \in \partial \mathcal{B}_{E_0}$ for all $t \in [t_{2k+1},t_{2k+2}]$ and for all $k \in \{0,\ldots,M\}$, and
	$e(t) \in \mathring{\mathcal{B}}_{E_0}$ for all $t \in (t_{2k},t_{2k+1})$ and for all $k \in \{1,\ldots,M\}$.
	Therefore, for all $t \geq t_1$,
	the trajectory tracking error $e(\cdot)$ is bounded to reside within $\overline{\mathcal{B}}_{E_0}$ and
	$\Vert \tilde{f}(t) \Vert \leq \Vert \tilde{f}(t_1) \Vert$ for all $t \geq t_1$.
	The case whereby 
	$e(t_0) \in \mathring{\mathcal{B}}_{E_0}$ can be deduced following similar arguments and is omitted for brevity.  
\end{proof}

Theorem \ref{theorem_DPS} provided a sufficient condition for the adaptive observer \eqref{eqn_state_estimator} and the adaptive law \eqref{eqn_adaptive_law_infinite_dim} to estimate the plant state.
At steady state, the observation error is commensurate with the known upper bound $\overline{\delta}$ on the measurement error $\delta(\cdot)$.
Ideally, if $\delta(t) \equiv 0_m$ for all $t \geq t_0$, then Theorem \ref{theorem_DPS} guarantees asymptotic convergence of the observation error to zero.

Theorem \ref{theorem_DPS} is based on the assumption of the uniqueness of the solutions of \eqref{eqn_plant_dynamics} and \eqref{eqn_state_estimator} for all initial conditions.
Sufficient conditions to verify this assumption can be deduced by proceeding as in the proof of Theorem 5.4 of \cite{RKHS_MRAC_book}, and are omitted for brevity.

\section{Implementable Adaptive Laws}\label{section_implementation}
The adaptive law \eqref{eqn_adaptive_law_infinite_dim} is a distributed parameter system (DPS), and, hence, is no implementable in practice.
In this section, we present finite-dimensional approximations of \eqref{eqn_adaptive_law_infinite_dim}.
However, as discussed in \cite[Ch. 5]{RKHS_MRAC_book} in the context of adaptive control systems design over native spaces, finite-dimensional approximations of DPSs induce error terms that must be accounted for by a robustification of the adaptive mechanism. 
In this section, we discuss a smooth dead-zone modification of the adaptive estimation system presented in Section \ref{section_Adaptive_Law}.
The classical dead-zone modification of adaptive laws \cite{1103112} is discontinuous and, hence, breaks the canonical assumptions of Lipschitz continuity of the tracking error dynamics.
Furthermore, in practice, the dead-zone modification of MRAC may induce chattering.
For these reasons, a smooth implementation of the dead-zone modification is considered.



\subsection{Smoothed Dead-Zone Modification}
Consider the \emph{finite-dimensional adaptive law}
\begin{align}
    \dot{\hat{f}}_N(t) &= \sigma_0(\normRd{e}(t),d,\varepsilon) \Gamma_f \Pi_N \Knl_{y(t)} (y(t)-C\hat{x}(t)), \nonumber \\
    &\hspace{9.0em} \hat{f}_N(t_0) = f_0, \quad t \geq t_0, \label{eqn_final_learning_laws}
\end{align}
where the \emph{dead-zone width} $d$ is such that
\begin{align}
    d &> d_N \nonumber \\
    &\triangleq \frac{2 \Vert C \Vert _2(\sup_{y\in \YY}\bm{\Pf}_N(y) \normvH{(I-\Pi_N)f} +  \Vert PL \Vert _2 \overline{\delta})}{\lambda_{\min}(W^{\rm T} W + \epsilon P)},\label{eqn_minimal_dead_zone}
\end{align}
where
$\lambda_{\min}(\cdot)$ denotes the eigenvalue of its argument with the smallest real part.
The function
$\sigma_0 : \RR \times \RR \times \RR \to \RR$ is a $\Delta$-admissible dead-zone in its first argument.

\begin{defn}[{\cite{boffi2022nonparametric}}]
    Let $\Delta>0$. A continuously differentiable function $\sigma_\Delta:\RR^+\to \RR$ is an $\Delta$-\emph{admissible dead-zone} if
    \begin{itemize}
        \item $0\leq \sigma_\Delta$ and $\sigma_\Delta(x)=0$ for all $x\in[0,\Delta]$,
        \item $0\leq \dot{\sigma}_\Delta$ and $\sigma_\Delta(x)=0$ for all $x\in[0,\Delta]$,
        \item $\dot{\sigma}_\Delta(\cdot)$ is locally Lipschitz,
    \end{itemize}
\end{defn}

An example of an admissible dead-zone is given by
\begin{align}
    &\sigma_0(\normRd{e},d,\varepsilon) \nonumber \\
     &\triangleq 
	 \left\{
	\begin{array}{ll}
		0,  & \mbox{if } \normRd{e} \leq d, \\
		\dfrac{(\normRd{e} - d)^2}{2\varepsilon},  & \mbox{if } \normRd{e} \in [d,d+\varepsilon], \\
		\normRd{e} - (d + \varepsilon),  & \mbox{if } \normRd{e} \geq d+\varepsilon, \\
	\end{array} \right. \label{eqn_ch10_function_mu_dead-zone_int}
\end{align}
where the \emph{dead-zone width} $d>0$ and
the \emph{buffer zone width} $\varepsilon>0$ are user-defined. The first derivative of this function is given by
\begin{align}
	 \dot{\sigma}_0(\normRd{e},d,\varepsilon) = 
	 \left\{
	\begin{array}{ll}
		0,  & \mbox{if } \normRd{e} \leq d, \\
		\dfrac{\normRd{e} - d}{\varepsilon},  & \mbox{if } \normRd{e} \in [d,d+\varepsilon], \\
		1,  & \mbox{if } \normRd{e} \geq d+\varepsilon, \\
	\end{array} \right. \label{eqn_ch10_function_mu_dead-zone}
\end{align}
which is Lipschitz continuous.

\begin{theorem}\label{theorem_deadzone_finite_dimensional}
    Consider the plant model \eqref{eqn_plant_dynamics}, the adaptive observer \ref{eqn_state_estimator}, and the adaptive law \eqref{eqn_final_learning_laws}.
    If the dead-zone $d$ width satisfies the condition in \eqref{eqn_minimal_dead_zone}, then there exists $T\in [t_0,\infty)$ such that $\normRd{e(t)} \triangleq \normRd{x(t)-\hat{x}(t)}\leq d$ for all $t>T$.
\end{theorem}

\begin{Proof}
    The proof follows by proceeding as in the proof of Theorem 5.5 in \cite{RKHS_MRAC_book}, and is omitted for brevity. 
    $\hfill$
\end{Proof}

The ultimate bound postulated by Theorem \ref{theorem_deadzone_finite_dimensional} correlates the observation error to the ability of capturing matched uncertainties through the native space approximation $\vH_N$.
Indeed, it follows from \eqref{eqn_minimal_dead_zone} that the ultimate bound on the observation error varies with the ability of $\vH_N$ to capture $\vH$, which is measured by the power function \eqref{eqn_power_function}.
As discussed in \cite[Ch. 3]{RKHS_MRAC_book}, the denser the distribution of kernel centers $\Xi_N$, the smaller the power function, and, hence, the smaller the width of the dead-zone the user can enforce.
However, the larger the number of centers, the larger the computational cost.
Indeed, the dimension of the Grammian matrix grows with the square of kernel centers, that is, it grows proportionally to $N^2$.
The ultimate bound on the estimation error also grows with $\Vert (I - \Pi_N) f \Vert_{\vH}$, which is an alternative measure of the ability of $\vH_N$ to approximate $\vH$.
Finally, as expected, the observation error grows with the upper bound $\overline{\delta}$ on the observation disturbance.


\subsection{Coordinate Implementation}
To implement the finite-dimensional adaptive law \eqref{eqn_final_learning_laws}, we let
\begin{align}
    \hat{f}_N(t,\cdot)=\sum_{j=1}^{N}\Knl_{\xi_j}(\cdot) \hat{\alpha}_j, \quad t \geq t_0,
\end{align}
where $\Xi_N \triangleq \{\xi_1, \ldots, \xi_N \}$ denotes the set of centers.
Thus, we observe that, for all $t \in [t_0,\infty))$,
\begin{align}
    &\sum_{j=1}^{N}\Knl_{\xi_j}(\cdot) \dot{\hat{\alpha}}_j(t) \nonumber \\
    &\quad = \Gamma_f \Pi_N \Knl_{y(t)}(\cdot) (y(t) - C \hat{x}(t)) \nonumber\\
    &\quad = \Gamma_f \ipvH{\Knl_{\xi_i}(\cdot)}{\Pi_N \Knl_{y(t)}(\cdot)} (y(t) -C\hat{x}(t)) \nonumber\\ 
    &\quad = \Gamma_f \ipvH{\Pi_N \Knl_{\xi_i}(\cdot)}{ \Knl_{y(t)}(\cdot)} (y(t)-C\hat{x}(t)) \nonumber\\
    &\quad = \Gamma_f \ipvH{\Knl_{\xi_i}(\cdot)}{ \Knl_{y(t)}(\cdot)} (y(t)-C\hat{x}(t))\nonumber \\
    &\quad = \Gamma_f \Knl_{\xi_i}(y(t)) (y(t)-C\hat{x}(t)), \label{eqn_temp01}
\end{align}
and
\begin{align}
        \sum_{j=1}^{N}\Knl_{\xi_j}(\cdot) \dot{\hat{\alpha}}_j(t) &= \sum_{j=1}^{N}\ipvH{\Knl_{\xi_i}(\cdot)}{\Knl_{\xi_j}(\cdot)} \dot{\hat{\alpha}}_j(t) \nonumber \\
        &\quad = \sum_{j=1}^{N}\Knl(\xi_i,\xi_j) \dot{\hat{\alpha}}_j(t) \nonumber\\
        &\quad = \sum_{j=1}^{N}\KK_{\Xi_N,ij} \dot{\hat{\alpha}}_j(t). \label{eqn_temp02}
\end{align}
Thus, it follows from \eqref{eqn_temp01} and \eqref{eqn_temp02} that
\begin{align}
    \dot{\hat{\alpha}}(t) &= \Gamma_f \KK^{-1}_{\Xi_N} \Knl_{\Xi_N}(y(t))(y(t) -C\hat{x}(t)), \nonumber \\
    &\hspace{9.0em} \hat{\alpha}(t_0) = \alpha_0, \quad t \geq t_0,
\end{align}
where $\KK_{\Xi_N}$ is given by \eqref{eqn_grammian_matrix}, and
$\Knl_{\Xi_N}(\cdot)$ is given by \eqref{eqn_kernel_matrix}.


\section{Numerical Example}\label{section_numerical_example}
In this section, we apply the proposed theoretical framework to the problem of estimating the state of a rigid body.
First, we will address the problem of estimating its position from the measured translational velocity.
Thus, we will address the problem of estimating its angular position.

\subsection{Estimating the Translational Position}
The translational dynamics of a rigid body are captured by \eqref{eqn_plant_dynamics_state} with
$x(t) = \left[r^{\rm T}(t), v^{\rm T}(t) \right]^{\rm T}$, $t \geq t_0$,
$r : [t_0,\infty) \to \RR^3$ denoting the position of the body's center of mass in an inertial reference frame $\mathbb{I}$,
$v : [t_0,\infty) \to \RR^3$ denoting the velocity of the body's center of mass in $\mathbb{I}$,
$A = \begin{bmatrix}
    0_{3 \times 3} & I_3 \\ 0_{3 \times 3} & 0_{3 \times 3}
\end{bmatrix}$,
$B = \begin{bmatrix}
    0_{3 \times 3}& m^{-1} I_3 
\end{bmatrix}^{\rm T}$,
$m > 0$ denoting the mass of the body,
$f \in \vH$ denoting some force acting on the body, and
$\xi : [t_0,\infty) \to \RR^3$ capturing some force that does not lie in the user-defined native space $\vH$.
In problems of practical interest, the translational position is deduced from the translational velocity, which is measured by fusing data from inertial measurement units (IMUs) and other sources of information, such as cameras.
Thus, in general \eqref{eqn_plant_dynamics_output} captures the measured output with $C = \left[0_{3 \times 3}, I_3 \right]$.
Furthermore, it is worthwhile noting how, in general, external forces acting on the body are functions of its velocity and, hence, a representation of the matched uncertainty in \eqref{eqn_plant_dynamics_state} as $E_{y(t)} f$, where $y(t) = v(t) + \delta(t)$, is adequate;
fluid dynamic forces, for instance, verify this assumption. 
The measurement error $\delta(\cdot)$ captures the unavoidable noise of the sensors.

\subsection{Estimating the Angular Position}
The rotational equations of motion of a rigid body are given by
\begin{subequations}\label{eqn_rotational_eom}
    \begin{align}
        \dot{\eta}(t) &= J(\eta(t)) \omega(t), \quad \eta(t_0) = \eta_0, \quad t \geq t_0, \label{eqn_rotational_eom_kin} \\
        \dot{\omega}(t) &= I^{-1} \left[u(t) - \omega^{\times}(t) I \omega(t) + E_{y(t)} f \right] + \xi(t), \nonumber \\
        &\hspace{12.0em} \omega(t_0) = \omega_0, \label{eqn_rotational_eom_dyn}
    \end{align}
\end{subequations}
where
$\eta(t) \triangleq [\phi(t),\theta(t), \psi(t)]^{\rm T} \in \mathcal{D}$  denotes the \emph{AUV attitude} through a 3-2-1 sequence of implicit Euler angles \cite[Ch. 1]{l2017mathematical} of the body reference frame $\mathbb{J}(\cdot)$ relative to $\mathbb{I}$,
$\mathcal{D} \triangleq [0,2\pi) \times \left(- \frac{\pi}{2}, \frac{\pi}{2} \right) \times [0,2\pi)$,
$\omega : [t_0,\infty) \to \RR^3$ denotes the \emph{angular velocity},
$(\cdot)^{\times} : \RR^3 \to {\rm so}(3)$ denotes the \emph{cross-product operator},
$I \in \RR^3$ denotes the symmetric, positive-definite inertia matrix,
$f \in \vH$ denotes moments of external forces acting on the body, and
$\xi : [t_0,\infty) \to \RR^3$ captures moments of forces that can not be modeled as elements of a native space.

It is well-known that \eqref{eqn_rotational_eom} can not be reduced to the same form as \eqref{eqn_plant_dynamics_state}.
However, \eqref{eqn_rotational_eom_dyn} is in the same form as \eqref{eqn_plant_dynamics_state} with
$x(t) = \omega(t)$, $t \geq t_0$,
$A = 0_{3 \times 3}$, and
$B = I^{-1}$.
In problems of practical interest, it is common practice to directly measure the angular velocity using IMUs so that $C = I_3$.
Thus, we apply the proposed framework to deduce $\omega(\cdot)$ despite uncertainties due to $f$, $\xi(\cdot)$, and $\delta(\cdot)$.

\subsection{Numerical Results}
To implement the proposed results, we consider the Sobolev-Matérn kernel $\Knl(\xi,x)= \sigma( \Vert \xi - x \Vert ) I_3$, where 
\begin{align}
      \sigma(r) = \frac{2\pi^{d/2}}{\Gamma(k)}{K}_{k-d/2}(r/2)^{k-d/2}, \quad r \geq 0
\end{align}
for $d,k\in \NN$.
Additional details of this kernel function can be found in \cite[Ch. 3]{RKHS_MRAC_book}.
The adaptive rate for both examples is set $\Gamma_f=1$.

To estimate the translational position of the rigid body, let $m = 4$ kg and
$u(t) =-K_{P,r}(x(t)-x_r(t)) + B^{\dagger} \dot{x}_r(t)$ where
$K_{P,r} = \begin{bmatrix}
    I_{3 \times 3} &I_{3 \times 3} 
\end{bmatrix}$
and
$x_r(t) \triangleq \begin{bmatrix}
    \sin(t) & \cos(t)& \sin(2t) & \cos(t) &-\sin(t) &2\cos(2t)
\end{bmatrix}^{\rm T}$
denotes the user-defined reference trajectory.
We set the deadzone $d = 0.01$ and buffer zone $ \varepsilon = 0.01$.
The nonlinear external force $f$ was chosen as
$f = \left[\cos(x_4^2), \sin(x_5^2) +\sin(x_4), \cos(x_6)+\sin(x_5) \right]^{\rm T}$, and
the disturbance $\delta(t)=0.008(\sin(0.5t)+\cos(0.5t))$.
The basis centers for this example are placed on a three-dimensional lattice in the velocity space with three centers on each dimension, totaling 27 basis centers.
The range of the 3D lattice is $[-1, 1]$ for all three dimensions.

To estimate the attitude of the rigid body, the deadzone was chosen to be $ d=0.1$ and the buffer zone was $\varepsilon=0.01$.
The inertia matrix considered in this example is $I=\text{diag}[0.2,15,15]$ kg m$^2$.
The control input is set 
$u(t) = -K_{P,\omega}(\omega(t) - \omega_r(t)) + \dot{\omega}_r(t)$, where $K_{P,\omega}=10\cdot I_3$.
The reference trajectory is chosen as
$\omega_r(t) = \left[0.1\cos(0.1t), 0.1\sin(0.1t), 0.1\tanh(0.1t) \right]^{\rm T}$, $t \geq t_0$.
In addition to the Coriolis term $\omega^{\times}(t) I \omega(t)$, $t \geq t_0$, the matched uncertainty in the rotational dynamics was chosen to be $f=-0.001\|\omega\|\omega$.
The measured output disturbance is chosen as $\delta(t)=0.05\sin(5t)$, $t \geq t_0$.
The basis centers for this example are placed in the space of angular velocities in the same manner as the transitional dynamics example.
The range of the 3D lattice is $[-0.1, 0.1]$ for all three dimensions.

\begin{figure}
    \centering
    \includegraphics[width=0.82\linewidth]{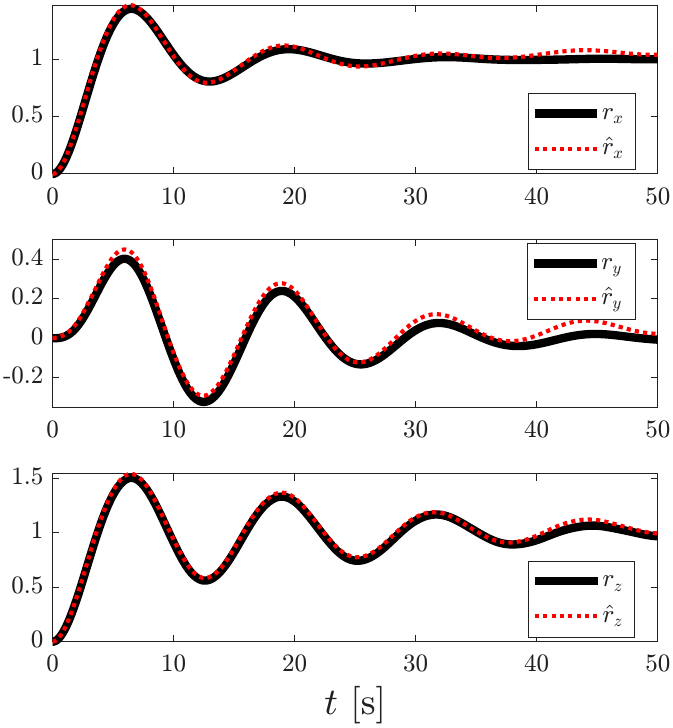}
    \caption{Actual translational position and the estimated translational position as functions of time}
    \label{fig_position}
\end{figure}

\begin{figure}
    \centering
    \includegraphics[width=0.8\linewidth]{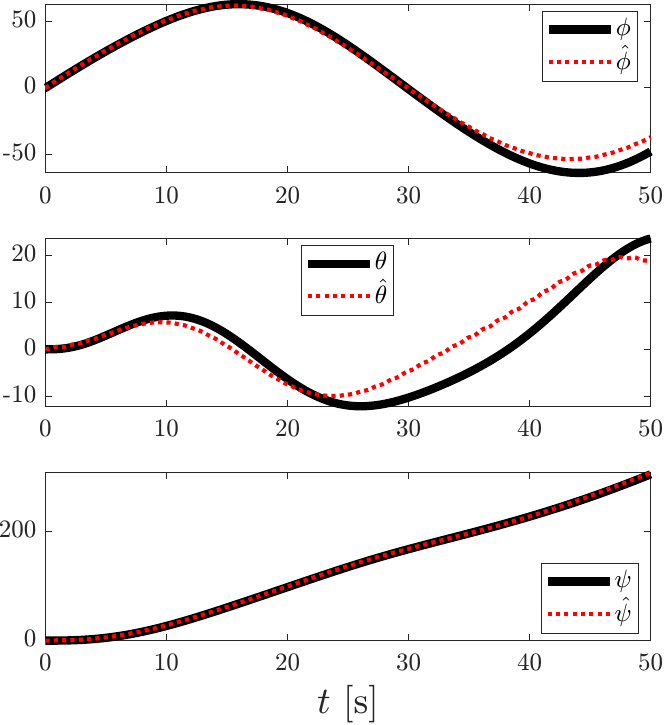}
    \caption{Actual angular position and estimated angular position as functions of time}
    \label{fig_Rotation_Angle}
\end{figure}

\begin{figure}
    \centering
    \includegraphics[width=0.8\linewidth]{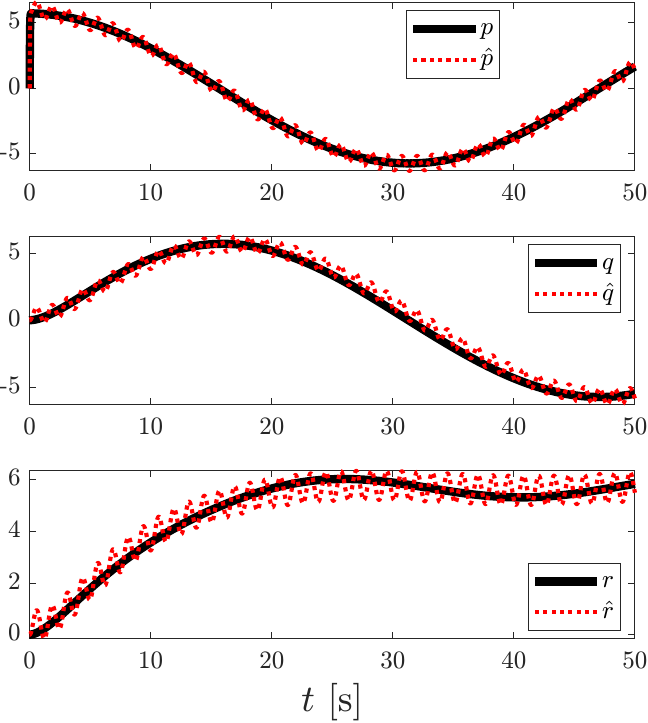}
    \caption{Actual angular rate and estimated angular rate as functions of time}
    \label{fig_Rotation_Rate}
\end{figure}

Figures \ref{fig_position}--\ref{fig_Rotation_Rate} capture the main results of the proposed simulations.
Figure \ref{fig_Rotation_Angle} shows a sizable error in the pitch angle and the roll angle estimation.
This error is the result of a large deadzone, which is designed to cover both $\delta(t)$ and the estimation error. This behavior highlights the limitation of the smoothed deadzone.
To prevent parameter drift, the estimation performance must be sacrificed. 

\section{Conclusion}\label{section_conclusion}
This paper presented an adaptive observation framework for uncertain nonlinear systems by embedding unknown dynamics into a vector-valued reproducing kernel Hilbert space.
The method extends adaptive estimation to both infinite- and finite-dimensional adaptive laws, incorporating a smoothed dead-zone modification to ensure boundedness and robustness.
Numerical results concerning the dynamics of rigid bodies demonstrated the applicability of the proposed theory. 

Future work directions are multi-fold. 
Of great interest will be the addition of non-deterministic matched, unmatched, and measurement disturbances.
Additional areas of interest concern merging adaptive control and estimation techniques over native spaces for an integrated output-feedback MRAC framework.

\bibliographystyle{IEEEtran}
\bibliography{reference}
\end{document}